\documentclass[10pt,conference]{IEEEtran}
\IEEEoverridecommandlockouts
\usepackage{cite}
\usepackage{amsmath,amssymb,amsfonts}
\usepackage{algorithmic}
\usepackage{graphicx}
\usepackage{textcomp}
\usepackage{xcolor}
\usepackage[hyphens]{url} 
\def\BibTeX{{\rm B\kern-.05em{\sc i\kern-.025em b}\kern-.08em
    T\kern-.1667em\lower.7ex\hbox{E}\kern-.125emX}}
\begin{document}

\IEEEoverridecommandlockouts\IEEEpubid{\makebox[\columnwidth]{ 978-1-6654-3540-6/22~\copyright~2022 IEEE \hfill} \hspace{\columnsep}\makebox[\columnwidth]{ }}

\title{Taming Hybrid-Cloud Fast and Scalable Graph Analytics at Twitter\\
}

\author{\IEEEauthorblockN{Chunxu Tang\textsuperscript{\textsection}, Yao Li\textsuperscript{\textsection}, Zhenxiao Luo, Mainak Ghosh, Huijun Wu, Lu Zhang, Anneliese Lu, Ruchin Kabra,\\Nikhil Kantibhai Navadiya, Prachi Mishra, Prateek Mukhedkar and Vrushali Channapattan}
\IEEEauthorblockA{
\textit{Twitter, Inc.}\\
San Francisco, California, USA \\
\{chunxut, yaoli, zluo, mghosh, huijunw, luz, anneliesel, rkabra, nnavadiya, prachim, pmukhedkar, vrushali\}@twitter.com}
}

\maketitle
\begingroup\renewcommand\thefootnote{\textsection}
\footnotetext{The authors contribute equally to this work.}
\endgroup

\begin{abstract}
We have witnessed a boosted demand for graph analytics at Twitter in recent years, and graph analytics has become one of the key parts of Twitter's large-scale data analytics and machine learning for driving engagement, serving the most relevant content, and promoting healthier conversations. However, infrastructure for graph analytics has historically not been an area of investment at Twitter, resulting in a long timeline and huge engineering effort for each project to deal with graphs at the Twitter scale. How do we build a unified graph analytics user experience to fulfill modern data analytics on various graph scales spanning from thousands to hundreds of billions of vertices and edges?

To bring fast and scalable graph analytics capability into production, we investigate the challenges we are facing in large-scale graph analytics at Twitter and propose a unified graph analytics platform for efficient, scalable, and reliable graph analytics across on-premises and cloud to fulfill the requirements of diverse graph use cases and challenging scales. We also conduct quantitative benchmarking on Twitter's production-level graph use cases between popular graph analytics frameworks to certify our solution.
\end{abstract}


\section{Introduction}

As one of the largest social media globally, Twitter manipulates multiple large-scale graphs with billions of vertices and edges. Implementing graph analytics on such large-scale graphs has been a long-lasting pain. Moreover, we have witnessed a growing need for large-scale graph analytics at Twitter in recent years. For example, recommendation teams are running PageRank and topic similarity tasks on the user-follow graph to measure the influence of Twitter users. Health teams are running multi-account detection on the safety graph to detect any or all Twitter accounts owned by the same person and combined connected users jobs to build relationships between users.

These demands lead graph analytics to be a strategic bet for Twitter as part of the Unified Data and Machine Learning initiative. Our thesis with the bet is that by capturing the complexity and richness of graph analytics, engineering teams will be able to use graph data assets, for example, Twitter user-follow graph, to unlock a better understanding of the Twitter ecosystem. Hence, users will experience a safer, more personalized, and more relevant Twitter, eventually leading to better user retention. 

Today, a typical Twitter's Graph ML (machine learning) project involves as much as six months of effort to build the initial dataset, creating a huge barrier to iterative development. Right now, most of these data preparation, ingestion, and processing pipelines are ad-hoc Scalding \cite{Scalding} jobs. Making changes to these jobs is costly and time-consuming, which limits the capability to fully explore and leverage graph data in ML applications. There are two primary reasons for this long timeline: 1) no flexible graph infrastructure exists today at Twitter, and hence the teams need to create bespoke data pipelines and querying solutions for each project, 2) working with Twitter large graphs, which cannot be fit in the memory of a single machine, is extremely time-consuming that stops us from fast iteration. In order to resolve these, we propose a unified graph analytics platform towards fast and scaling analytics capability for retrieval of relationships between Twitter's nouns (users, tweets, etc.) based on a series of quantitative experiments on Twitter's large-scale production-level graphs. Specifically, this paper makes the following contributions:

\begin{enumerate}
    \item We share the challenges and lessons learned from historically developing and operating large-scale graph analytics at Twitter.
    \item Motivated by challenges we identified in large-scale industrial graph analytics, we propose a hybrid-cloud graph analytics infrastructure which
    \begin{itemize}
        \item Provides production support for efficient, scalable, and reliable graph analytics frameworks to run graph analytics and reduce the iteration time of Graph ML.
        \item Provides production support for graph libraries of common graph algorithms and operations so that users do not need to reinvent the wheel, such as PageRank and combined connected components.
        \item Supports graph analytics of large scale, increasing the supported graph size to the scale of tens of billions of vertices and edges.
    \end{itemize} 
    \item We benchmarked some popular graph systems across graph processing systems and graph databases on enterprise-grade large-scale graphs with billions of vertices and edges under multiple production use cases.
\end{enumerate} 

\section{Challenges}
\label{sec.challenges}

From our development and operational experience in large-scale graph analytics, we observe a series of challenges in establishing a modern large-scale graph analytics platform, ranging from graph variety and scale, performance, and development to user-friendliness and cross-environments. Specifically, take Twitter as an example, we notice the following existing challenges.

\subsection{Complicated Graph Types and Scales at Twitter}

\begin{figure}[htbp]
\centerline{\includegraphics[width=0.45\textwidth]{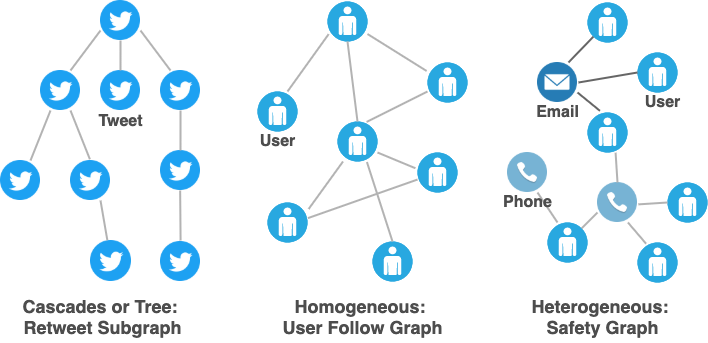}}
\caption{Graph types at Twitter.}
\label{fig.graph_types}
\end{figure}

There are various graph types at Twitter, as shown in Figure \ref{fig.graph_types}, that scale from thousands to tens of billions of vertices or edges. Typically, we can categorize them into three types:

\begin{itemize}
\item Cascades or Tree: This kind of graphs describe the cascading relationships, such as retweet and knowledge graphs. They usually have clear structures like a tree and contain fewer vertices (thousands of). 
\item Homogeneous graphs: Homogeneous graphs contain a single type of vertices and edges, such as the user-follow graph which describes the follow relationship among Twitter users. One large graph may contain millions of vertices and billions of edges. We often observe small-world structures  in such type of graphs.
\item Heterogeneous graphs: From homogeneous graphs, more complexities can be involved by adding multiple types of vertices and edges with all kinds of properties in a single large graph. As a result, vertex and edge count can rise to billions or even tens of billions, and the graph structure also becomes unpredictable depending on concrete use cases. 
\end{itemize}

Various graph types and scales usually correspond to different use cases that require diverse graph algorithms and query frequency. For example, small graphs or sub-graphs are often ad-hocly used for graph analytics experiments, while large heterogeneous graphs can give us more information for a holistic view of high-level community or anomaly detection. We need to accommodate the variety of requirements in Twitter within an inclusive graph analytics ecosystem.

\subsection{Long Running Time}

Based on our survey, 71.4$\%$ graph users are suffering from long-running time for their graph analytics use cases. Currently, most of the users are leveraging Scalding \cite{Scalding} (Scala library built on top of MapReduce \cite{dean2008mapreduce}) or federated SQL systems \cite{tang2021hybrid} based on query engines such as BigQuery \cite{BigQuery} and Presto \cite{sethi2019presto,luo2022batch}, to run graph analytics. Due to lack of graph indices, insufficient caching, and poor graph libraries support, graph analytics jobs easily run in tens of hours. For example, it takes more than 11 hours to complete one iteration of a PageRank job and around 20 hours to complete a combined connected users job. The huge job latency prevents users from conducting fast graph analytics. What is worse, to guarantee that graph analytics jobs are completed within one day, some teams have to run the jobs on a sample of data, resulting in sacrificing accuracy. 

\subsection{Lack of Graph Library Support}

Due to poor or no graph analytics libraries support, users have to reinvent the wheel repeatedly. For example, a number of teams are running PageRank and node similarity jobs, and each of them has its own implementation of PageRank or node similarity algorithms in Scalding or Python scripts based on BigQuery. It takes up to 6 person-months for development for each of the teams, and substantial engineering efforts are also needed to maintain the code, make changes, conduct result verification, etc., which is costly and time-consuming.

\subsection{Lack of User Interface}

Currently, there is no user interface, customer onboarding tools, or code templates for graph analytics. These are considerable barriers for users to onboard graph analytics jobs. Many users interested in graph analytics gave up after finding that the only way to run graph analytics jobs is by writing complicated Scalding jobs. We need an intuitive way for users to play with graphs.

\subsection{Limited Graph Analytics Capability on GCP}

FlockDB \cite{FlockDB} is Twitter's standard graph storage which stores the many graphs of relationships between Twitter's users. It is currently on-prem only, and analytics over the graph data requires dumping the associations to HDFS (Hadoop Distributed File System) and then running Hadoop jobs through Scalding. Although the graph snapshots in HDFS can be replicated to GCP (Google Cloud Platform), people still suffer from limited graph analytics capability on the cloud, both data-wise and graph processing capability-wise. Twitter is strategically moving to the cloud now, aka Partly Cloudy \cite{partlycloudy}, especially for graph ML to leverage the GCP ML ecosystem. It is a burning demand that graph analytics can be entirely done on GCP where the training sits.

\section{Towards Unified Graph Analytics}

\subsection{Unified Graph Analytics Abstractions}

\begin{figure}[htbp]
    \centerline{\includegraphics[width=0.48\textwidth]{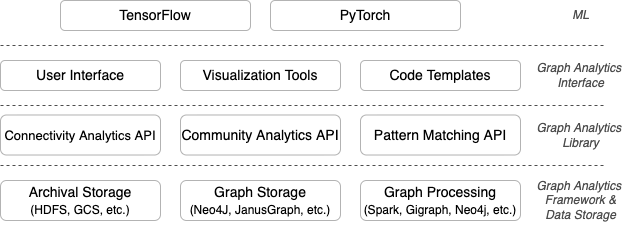}}
    \caption{Unified graph analytics stack.}
    \label{fig.layer}
\end{figure}

Figure \ref{fig.layer} depicts the logical building layers and blocks that constitute a unified graph analytics stack. On the bottom layer, graph analytics frameworks interact with data storage to provide graph processing capability in different environments at various scales. There are several well-known candidates for graph storage and processing that we can leverage, which will be discussed in the following subsection. On top of the frameworks, several graph analytics libraries corresponding to each underlying framework play an important role in providing the most efficient graph-related data manipulation capability and graph algorithm implementation. To enhance user experience, a graph analytics interface layer, which contains a unified user interface, visualization tools, and code templates for easy onboarding, is also required. This interface layer should be flexibly integrated with existing ML platforms for a smooth Graph ML experience and also be able to accommodate different underlying graph frameworks/databases and libraries with minimal effort.

\subsection{Graph Analytics Stack at Twitter}

\subsubsection{Graph Storage}

Based on an evaluation of multiple popular graph databases, we propose to adopt Neo4j in the unified graph analytics platform thanks to its extremely high popularity in graph databases \cite{graph-ranking} and high consistent query performance \cite{lissandrini2018beyond}. Neo4j \cite{miller2013graph} is a self-contained, full-featured graph database, offering additional functionality via graph algorithms and visualization libraries. It supports Cypher \cite{francis2018cypher}, a graph query language, natively. It can be deployed on GCP as an application in GKE and can scale accordingly, which complies with Twitter's Partly Cloudy strategy and provides flexible ad-hoc graph access like experimentation on GCP. Neo4j helps relieve the pain of building from open source software, configuring and maintaining the application ourselves, and free us from relying on or managing dependencies like Bigtable.

\subsubsection{Graph Processing} 

Spark is a unified analytics engine for large-scale data processing. It promises to be orders of magnitude faster than Hadoop MapReduce--especially for workloads that fit in cluster memory and have iterative algorithms. Compared to other candidates with similar scale capability, such as Giraph \cite{Giraph}, Spark, as one of the most widely-known distributed processing frameworks \cite{coimbra2021analysis}, wins out thanks to its interactive SQL-like interfaces, large open-source community, and deep integration with Scala (which is the primary programming language used at Twitter). 

Nowadays, modern graph databases such as Neo4j and JanusGraph \cite{JanusGraph} also support built-in graph analytics algorithms, especially for ad-hoc queries. Based on our evaluation illustrated in Section \ref{sec.evaluation}, we propose to leverage Spark to handle the most challenging large-scale graph analytics and Neo4j for small and medium-scale graphs, for example, those containing less than 100 million edges and vertices.

\subsection{Architectural Design}

\subsubsection{Graph Processing \& Storage}

\begin{figure}[htbp]
    \centerline{\includegraphics[width=0.45\textwidth]{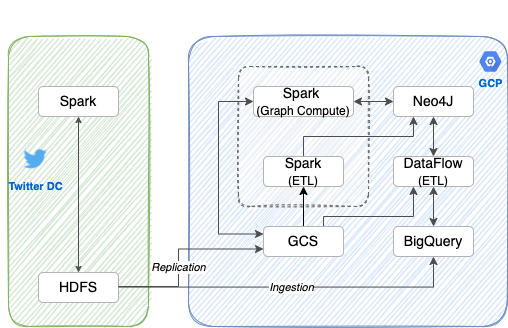}}
    \caption{High-level infrastructure for graph analytics at Twitter.}
    \label{fig.architecture}
\end{figure}

Figure \ref{fig.architecture} illustrates the high-level architectural design for the hybrid-cloud unified graph analytics platform at Twitter. The platform aims to serve a graph infrastructure both on-prem and on GCP that can scale up to support all use cases at Twitter to solve the challenges mentioned in Section \ref{sec.challenges}. The evaluation results in Section \ref{sec.evaluation} also help us make the judgment.

Spark serves as the distributed graph process framework for both on-premises and cloud environments. Neo4j, set up in the Google Cloud, also helps graph analytics on small and medium-scale graphs. It also stores graph data structures. Users could get the best speed of Neo4j for relatively small graphs using its Data Science Library, awesome user interface, and graph visualization.

\subsubsection{ETL Pipeline}

To connect all the pieces up and provide up-to-date graph data access, we need a configurable ETL (Extract, Transform, and Load) system that allows for flexible graph generation, graph algorithm execution, and results/queries serving either directly to consuming applications or storing intermediate results in BigQuery or GCS for further transformations. There are several flavors for the ETL pipelines:

\begin{itemize}
    \item For large-scale graph analytics relying on Spark, graph data will be retrieved from HDFS or GCS, and results will be persisted back to HDFS and GCS for down-streaming ML jobs. 
    \item Neo4j data ingestion and querying pipeline can either be deployed as a Spark application or realized in Beam/Dataflow. In addition, we can leverage the Neo4j-Spark-Connector APIs, which greatly ease development by using Spark DataFrame to create new nodes or return query results as DataFrame. Beam/Dataflow on GCP also has good support and extensive applications at Twitter and has been treated as the star product for next-generation general-purpose data processing.
\end{itemize}

\section{use case study and evaluation}
\label{sec.evaluation}


\subsection{Use Cases}

\subsubsection{Multi-Account Detection}

\begin{figure}[htbp]
\centerline{\includegraphics[width=0.3\textwidth]{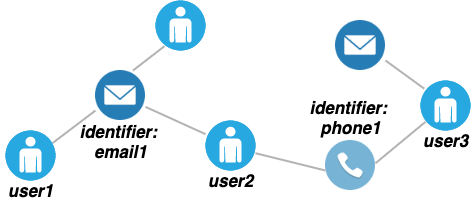}}
\caption{An example subgraph for multi-account detection.}
\label{fig.subgraph}
\end{figure}

The multiple account detection job aims to find all Twitter accounts owned by the same person. It is one of the key jobs for facilitating downstream ML to enhance Twitter health and safety. Two users are identified as the same user when an identifier directly connects them. For example, in Figure \ref{fig.subgraph}, email and phone serve as the identifiers. Hence \textit{user1} and \textit{user2} are the same user, \textit{user2} and \textit{user3} are the same user too, but \textit{user1} and \textit{user3} are not the same user in this case because they are not directly connected by an identifier. This is essentially a two-hop traversal problem: starting from each user and finding all the two-hop neighbors. This job runs upon four daily snapshot datasets containing 14.89 billion vertices and 30.86 billion edges of a heterogeneous graph.

\subsubsection{Combined Connected Users (Components)}

The problem of finding connected users is one step further of the multi-account detection, where the user relationship is transitive, i.e., if \textit{user1} and \textit{user2} are the same user, \textit{user2} and \textit{user3} are the same user, then \textit{user1} and \textit{user3} will be the same user. This job runs upon 2 daily snapshot datasets containing 2.41 billion vertices and 1.50 billion edges in total.

\subsection{Spark vs. Neo4j on Scalability}

Nowadays, both Spark (GraphFrame) and Neo4j support various popular graph analytics algorithms. Considering the very large-scale graphs in the social media industry, what options should we offer to customers to fit various graph scales? Can a single graph database cover all enterprise-grade graph analytics scenarios? We explore the scalability comparison between Spark and Neo4j on the combined connected Users job at Twitter.

We conduct the evaluation by gradually increasing the number of sampled nodes from the complete daily snapshot datasets. The output is a list of user IDs and corresponding component IDs. We set up an enterprise version Neo4j instance with 32 CPU cores and 128GB RAM. To make the compute resource similar, we run the Spark jobs with 32 executors, each with 1 CPU core and 4GB RAM.

\begin{figure}[htbp]
    \centerline{\includegraphics[width=0.45\textwidth]{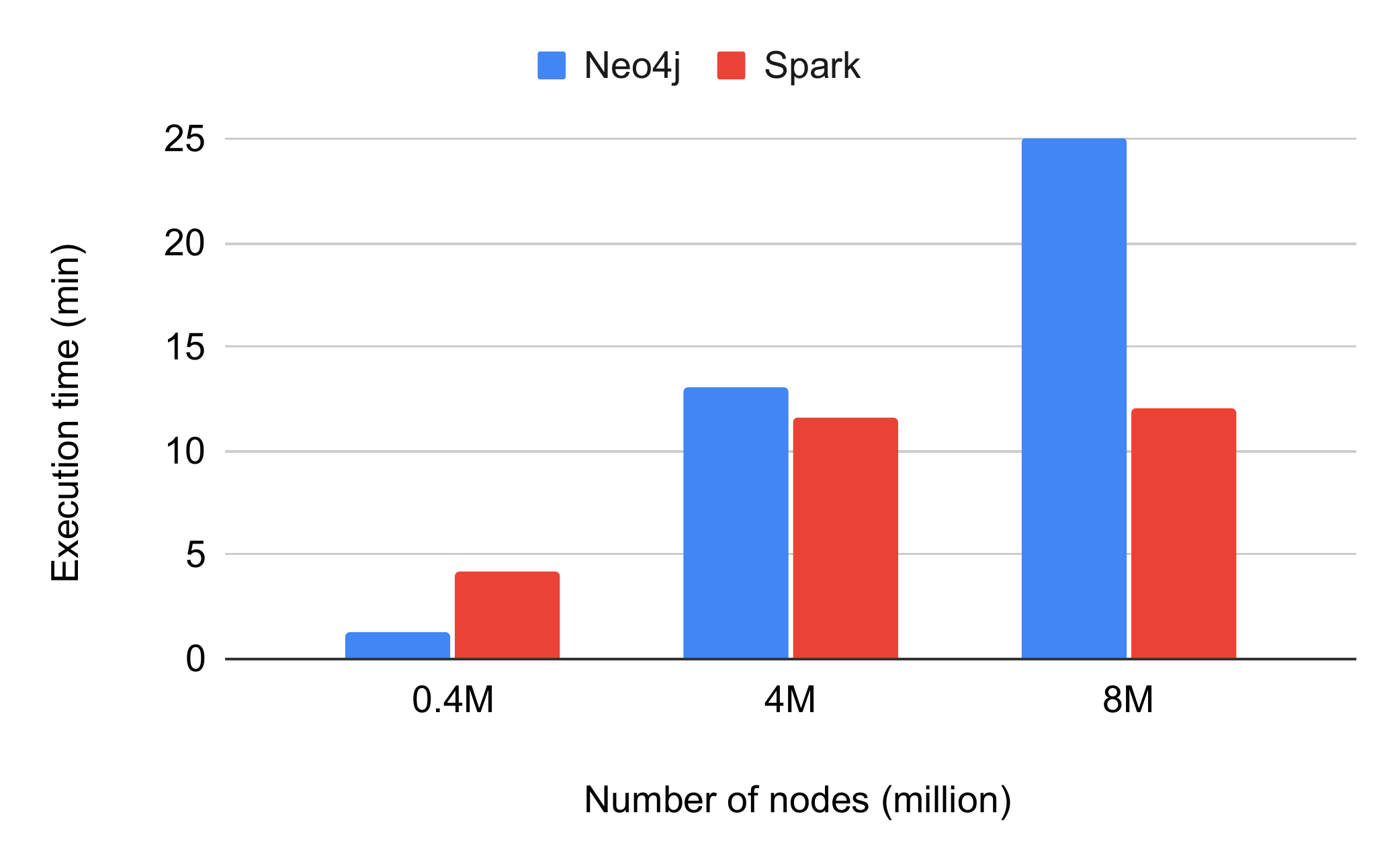}}
    \caption{Comparison of Spark and Neo4j on sampled combined connected users.}
    \label{fig.spark-neo4j}
\end{figure}

From our evaluation, when the graph scale is low, such as consisting of 0.4 million vertices, Neo4j outperforms Spark in running the combined connected users job. When the graph scale becomes higher such as consisting of around 10 million vertices, Spark is more efficient than Neo4j. But Spark itself also cannot cover all use cases. If we change the job output from a list of user IDs and component IDs to just the number of connected components, Neo4j dramatically outperforms Spark when there are around 10 million vertices. Neo4j takes less than 2 seconds to return the count, but Spark spends around 10 minutes. From our observation, Spark outperforms when the graph scale is very large or the output is very large, and Neo4j is more efficient when the graph scale is medium or only a small amount of output, such as a count or a limited number of rows, is required.

Overall, the comparison findings prove our architectural design that a single graph system cannot cover all industrial graph analytics scenarios, and a hybrid architecture combining graph processing systems and graph databases is required in modern enterprise-grade graph analytics.

\subsection{Spark on Large-Scale Graph Analytics}

We want to answer whether Spark with GraphFrame can handle the most challenging large-scale graphs at Twitter. We conducted experiments on Spark for both multi-account detection and combined connected users.

\subsubsection{Multi-Account Detection}

Currently, this production job is implemented in Scalding and running in the Hadoop cluster. It basically takes three steps: 1) find all the identifier neighbors for each user, 2) find all the user neighbors for each identifier, 3) join the result in steps 1) and 2), then group by the user. In order to limit the computation scale and reduce job running time, an extra \textit{MaxAdjacentNodes} restriction is added, which limits the number of neighbors of any vertex to no larger than 100. Under such a restriction, the edge count can be reduced to 22.29 billion. It takes 2-3 hours to generate the graph in the adjacency list representation and 2-3 hours to conduct the two-hop traversal, resulting in 4-6 hours in total for each day's data.

To solve the multi-account detection problem, we leverage the GraphFrames Motif Finding API, which can be used for searching for structural patterns in a graph. Specifically, the pattern we use is:

 \textit{(user1) $-$ [edge1]  $\rightarrow$ (identifier) $-$ [edge2] $\rightarrow$ (user2)}

\noindent where \textit{()} represents graph vertices and \textit{[]} represents graph edges.
We build the graph in GraphFrames directly from the original datasets and run the Motif Finding to find the two-hop searching results of the same users. We run the Spark job with 2 executors, each having 1 CPU core and 16GB RAM. 

\begin{figure}[htbp]
\centerline{\includegraphics[width=0.5\textwidth]{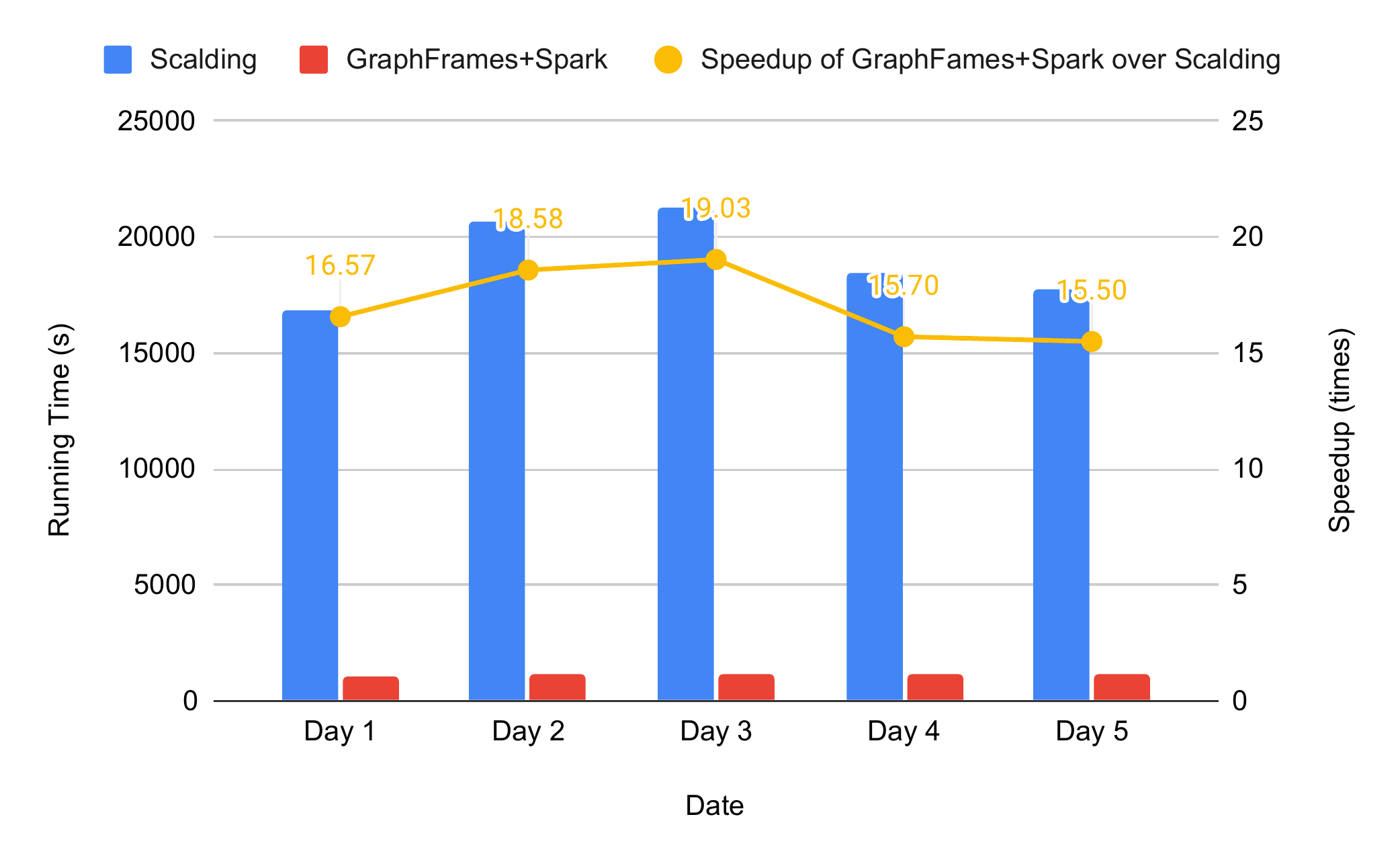}}
\caption{Multi-account detection job running time.}
\label{fig.multi_account_detection}
\end{figure}

\textbf{Efficiency.} As shown in Figure \ref{fig.multi_account_detection}, it takes around 20 minutes to finish running the multi-account detection job implemented in GraphFrames. Compared with that of the legacy Scalding jobs, it achieves around 17x speedup.

\textbf{Scalability.} As mentioned above, a \textit{MaxAdjacentNodes} restriction is added in the Scalding job to limit the scale. The relationship between \textit{MaxAdjacentNodes} and edge data lost percentage is shown in Table \ref{tab.edge_count_MaxAdjacentNodes}.

\begin{table}[htbp]
\caption{Data Lost Percentage vs. MaxAdjacentNodes}
\begin{center}
\begin{tabular}{|c|c|c|}
\hline
\textbf{\textit{MaxAdjacentNodes}} & \textbf{Edge Count} & \textbf{Lost Percentage ($\%$)}\\
\hline
10 & 20,695,731,797 & 32.9\\
100 & 22,290,700,453 & 27.8\\
1000 & 24,529,686,897 & 20.5\\
10000 & 25,809,948,026 & 16.4\\
100000 & 26,571,594,689 & 13.9\\
1000000 & 29,281,153,827 & 5.1\\
10000000 and larger & 30,859,836,432 & 0\\
\hline
\end{tabular}
\label{tab.edge_count_MaxAdjacentNodes}
\end{center}
\end{table}

In the Scalding job, \textit{MaxAdjacentNodes} is set to 100, which means 27.8$\%$ of edge data is actually lost, resulting in degradation in result accuracy. While in the GraphFrames+Spark solution, no \textit{MaxAdjacentNodes} restriction is required, so we can get accurate results in a shorter running time.

\textbf{Agility.} The development cycle of the multi-account detection job can be shortened from around 1 person-months to 0.25 person-months by using GraphFrames+Spark, which dramatically reduces engineering effort and iteration time.

\subsubsection{Combined Connected Users}

The legacy Twitter's Scalding combined connected users job has two steps: 1) find the connected components for each edge set, i.e., the edges connecting users and one identifier, 2) combine the connected components results of each edge set. It should be noted that there is no Connected Component API available in Scalding, so the job owner team actually implemented their own connected components algorithm with simplifications and trade-offs for runtime performance. As a result, it takes around 9-14 hours to find the connected components for each edge set and 12-15 hours to combine the results, resulting in 17-29 hours in total to finish the combined connected users job for each day's data. 

GraphFrames provides the Connected Components API. We build the graph in GraphFrames directly from the original datasets and call this API to get the results. We run the Spark job with 2 executors, each having 1 CPU core and 16GB RAM.

\textbf{Efficiency.} It takes around 40 minutes to finish running the combined connected users job implemented in GraphFrames. Compared with that of the legacy Scalding jobs, it achieves around 37x speedup.

\begin{figure}[htbp]
\centerline{\includegraphics[width=0.5\textwidth]{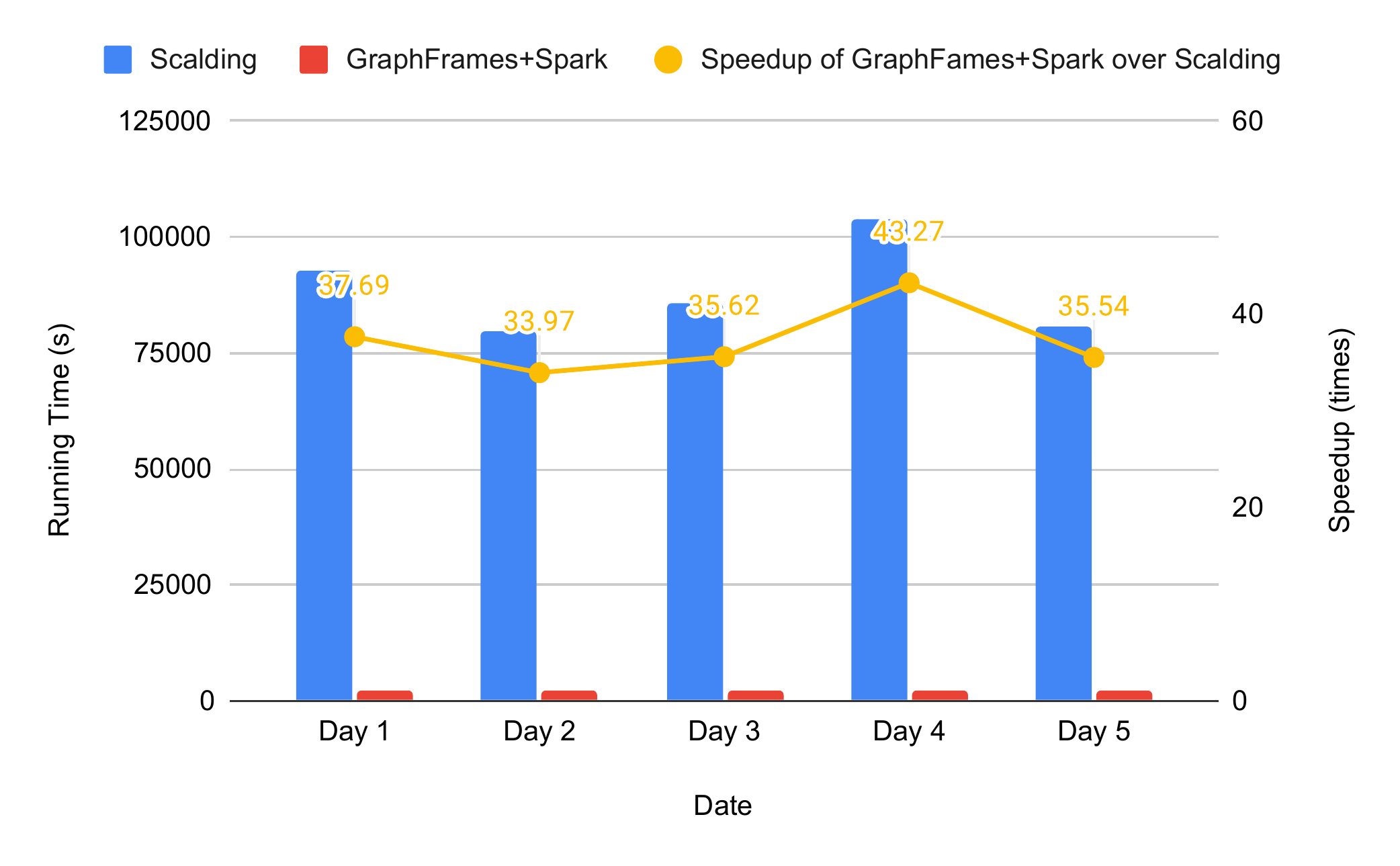}}
\caption{Combined connected users job running time.}
\label{fig.combined_connected_component}
\end{figure}

\textbf{Scalability.} The legacy Scalding jobs find the connected components separately for each identifier and then combine them together in another Scalding job. By contrast, the GraphFrames solution is more scalable by constructing a single graph that contains all the identifiers and edges and calculating the combined connected components directly upon this graph. Moreover, in terms of distinct user vertex count, the GraphFrames solution contains 72.4$\%$ more data than the Scalding one, resulting in more completed and accurate results.

\textbf{Agility.} By using GraphFrames and Spark, users do not have to reinvent the wheel for the connected components algorithm, and the development effort can be reduced from several person-months to 0.25 person-months.

\section{Related Work}

A large-scale graph analytics platform involves a wide spectrum of related domains. Here, we discuss related work in each domain.

\subsection{Graph Processing Framework and Systems}

In the recent decade, we witnessed a number of distributed graph processing systems developed, mostly targeting iterative batch-level graph processing. One of the pioneering works is Google's Pregel system \cite{malewicz2010pregel}, which employs a vertex-centric paradigm, based on the BSP (Bulk Synchronous Parallel) model \cite{gerbessiotis1994direct}. Motivated by Pregel, Giraph \cite{Giraph} is an open-sourced implementation of Pregel and has been deployed and extended in Meta (formerly Facebook) to process large-scale graphs with one trillion edges \cite{ching2015one}. GraphX \cite{gonzalez2014graphx}, built on top of RDDs (resilient distributed datasets), and GraphFrames \cite{dave2016graphframes}, built on top of dataframes, are both graph libraries that provide nice abstractions and dataflow optimizations for parallel graph processing on top of Apache Spark. 
In the proposed Twitter's unified graph analytics platform, we choose Spark with GraphFrames for graph processing thanks to interactive SQL-like interfaces, high ecosystem maturity, large open-source community, and deep integration with Scala.

\subsection{Graph Databases}

Graph databases store graphs based on entity relationships and support interactive queries with various graph traversal languages, such as Gremlin \cite{rodriguez2015gremlin} and Cypher \cite{francis2018cypher}. Gremlin, as a graph traversal machine and language, is widely supported in a large group of graph databases such as JanusGraph \cite{JanusGraph} and ComosDB \cite{reagan2018cosmos}. Gypher, as a declarative property graph language, was initially developed for Neo4j \cite{miller2013graph} and is now supported by other graph databases such as RedisGraph \cite{cailliau2019redisgraph}. Neo4j is a self-contained graph database, offering a rich set of graph algorithms and visualization libraries. It also has a data science library for more complicated algorithms and use cases such as fraud detection and link prediction. Similarly, JanusGraph, forked from TitanDB \cite{TitanDB}, is an open-sourced distributed graph database with flexible storage backends. In the proposed Twitter's unified graph analytics platform, we choose Neo4j as the graph database considering its extremely high popularity in graph databases \cite{graph-ranking}, high ecosystem maturity, and high and consistent query performance \cite{lissandrini2018beyond}.

\section{conclusion}
In order to provide fast and scalable graph analytics at Twitter, we designed a solution that leverages Spark and Neo4j for graph processing at large scales, with flexible ETL pipelines for graph generation and result retrieval for down-streaming ML jobs. Benchmarks on supported graph scale and performance of using GraphFrames library on top of Spark have been conducted, and results show impressive performance enhancement in terms of running time speedup - around 17x and 37x for the two typical large-scale graph ML feature extraction use cases and reliable support for tens of billion vertices and edges within a single graph. 

\section*{Acknowledgment}

We would like to express our gratitude to everyone who has served in the Twitter Graph Analytics Working Group for their profound insight and dedicated effort. Specially, we would like to thank Denis Garcia, Cindy McMullen, Rob Tandy, Yao Wu, Josh Whitney, Anna Wawrzyniak, Davide Eynard, Adam Crane, and Arash Aghevli for their extraordinary inputs. We also appreciate Cary Wun, Daniel Lipkin, Derek Lyon, and Srikanth Thiagarajan for their strategic vision, direction, and support to the team.

\bibliographystyle{IEEEtran}
\bibliography{library}

\end{document}